\documentclass[final,5p,times,twocolumn]{elsarticle}

\usepackage{amssymb}
\usepackage{lipsum}
\usepackage{amsthm}

\usepackage[colorlinks,allcolors={blue}]{hyperref}
\usepackage{amsmath}
\usepackage{natbib}
\usepackage{bm}

\usepackage{subfigure}
\usepackage{graphicx}
\usepackage{CJK}
\usepackage{xspace}  
\usepackage{multirow} 
\usepackage{dcolumn}
\usepackage{pstricks}
\usepackage[utf8]{inputenc}
\usepackage{psfrag}
\usepackage[T1]{fontenc}
\usepackage{booktabs}
\usepackage{url}
\usepackage[normalem]{ulem}    

\usepackage{lineno}

\newcommand{\ts}{\textsuperscript}
\newcommand{\tb}{\textsubscript}

\newcommand{\ghray}{$\gamma$-ray\xspace}
\newcommand{\grays}{$\gamma$ rays\xspace}

\newcommand{\mevu}{MeV/nucleon\xspace} 
 
\newcommand{\gcm}{g/cm\ts{2}\xspace}

\newcommand{\twoplus}{$2^+_1$\xspace}
\newcommand{\twotwoplus}{$2^+_2$\xspace}
\newcommand{\fourplus}{$4^+_1$\xspace}
\newcommand{\threeminus}{$3^-_1$\xspace}

\newcommand{\twotozero}{$2^+_1\rightarrow 0^+_{g.s}$\xspace}

\journal{Physics Letters B}

\begin{document}

\begin{frontmatter}



\title{Electromagnetic transition strength in $^{53}$Ca: the lifetime of the $5/2^-$ state}


            
\author[ayork]{S.~Chen}
\author[ayork]{M.~Petri}
\author[ayork]{S.~Paschalis}
\author[albnl]{H.~L.~Crawford}               
\author[ariken]{P.~Doornenbal}               
\author[ayork]{R.~Taniuchi}                  
\author[amadrid,aut,ariken]{K.~Wimmer}       

\author[afrib]{B.~A.~Brown}
\author[atriumf,amcgill]{J.~D.~Holt}
\author[atexas]{B.~S.~Hu}
\author[atsukuba]{T.~Miyagi}
\author[ariken,aut,ajaea]{T.~Otsuka}
\author[atsukuba]{Y.~Tsunoda}

\author[arcnp]{N.~Aoi}                       
\author[ariken]{H.~Baba}                     
\author[ayork]{M.~Bentley}
\author[ariken]{F.~Browne}                   
\author[albnl]{C.~Campbell}                  
\author[agsi]{Z.~Chen}                       
\author[ayork]{R.~Crane}
\author[aleuven]{H.~de~Witte}                
\author[albnl]{P.~Fallon}
\author[akoeln]{C.~Fransen}                  
\author[akoeln]{H.~Hess}                     
\author[arcnp]{E.~Ideguchi}                  
\author[arcnp]{S.~Iwazaki}                   
\author[aseoul]{J.~Kim}                      
\author[arcnp]{A.~Kohda}                     
\author[aut]{T.~Koiwai}                      
\author[ayork]{W.~Marshall}
\author[ariken]{B.~Mauss}                    
\author[aibs,ariken]{B.~Moon}                     
\author[asurrey]{T.~Parry}                   
\author[akoeln]{P.~Reiter}                   
\author[ariken,aut]{H.~Sakurai}              
\author[aut,aqnsi,ariken]{D.~Suzuki}                   
\author[ayork]{L.~Tetley}
\author[akoeln]{S.~Thiel}                    
\author[arcnp]{Y.~Yamamoto}                  

\address[ayork]{School of Physics, Engineering and Technology, University of York, Heslington, York YO10 5DD, UK}
\address[albnl]{Nuclear Science Division, Lawrence Berkeley National Laboratory, Berkeley, California 94720, USA}
\address[ariken]{RIKEN Nishina Center, 2-1 Hirosawa, Wako, Saitama 351-0198, Japan}
\address[amadrid]{Instituto de Estructura de la Materia, CSIC, E-28006 Madrid, Spain}
\address[aut]{Department of Physics, University of Tokyo, 7-3-1 Hongo, Bunkyo, Tokyo 113-0033, Japan}
\address[afrib]{Department of Physics and Astronomy and the Facility for Rare Isotope Beams, Michigan State University, East Lansing, MI 48824-1321, USA}
\address[atriumf]{TRIUMF 4004 Wesbrook Mall, Vancouver, British Columbia V6T 2A3, Canada}
\address[amcgill]{Department of Physics, McGill University, Montr\'eal, QC H3A 2T8, Canada}
\address[atexas]{Cyclotron Institute and Department of Physics and Astronomy, Texas A\&M University, College Station, Texas 77843, USA}
\address[atsukuba]{Center for Computational Sciences, University of Tsukuba, 1-1-1 Tennodai, Tsukuba 305-8577, Japan}
\address[ajaea]{Advanced Science Research Center, Japan Atomic Energy Agency, Tokai, Japan}
\address[arcnp]{Research Center for Nuclear Physics (RCNP), Osaka University, Mihogakoa, Ibaraki, Osaka 567-0047, Japan}
\address[agsi]{GSI Helmoltzzentrum f\"ur Schwerionenforschung GmbH, Planckstr. 1, 64291 Darmstadt, Germany}
\address[aleuven]{Institute for Nuclear and Radiation Physics, KU Leuven, Leuven B-3001, Belgium}
\address[akoeln]{Institut f\"ur Kernphysik, Universit\"at zu K\"oln, D-50937 Cologne, Germany}
\address[aseoul]{Department of Physics, Korea University, Seoul 02841, Republic of Korea}
\address[aibs]{Center for Exotic Nuclear Studies, Institute for Basic Science, Daejeon, 34126, Republic of Korea}
\address[asurrey]{Department of Physics, University of Surrey, Guildford GU2 7XH, UK}
\address[aqnsi]{Quark Nuclear Science Institute, the University of Tokyo, 7-3-1 Hongo, Bunkyo, Tokyo 113-0033, Japan}

\begin{abstract}
High-resolution in-beam $\gamma$-ray spectroscopy of \ts{53}Ca was performed 
at the Radioactive Isotope Beam Factory (RIBF) at RIKEN. 
Excited states in \ts{53}Ca were populated via one-neutron and one-proton removal reactions
from a \ts{55}Sc radioactive beam, and the emitted $\gamma$ rays were detected using
the HiCARI hybrid germanium detector array.
The lifetime of the first excited $5/2^-$ state in $^{53}$Ca was measured for the first time
through an analysis of Doppler-broadened $\gamma$-ray line shapes arising from relativistic recoil velocities.
A lifetime of $\tau = 11.3_{-11.3}^{+5.3\text{(stat)}+3\text{(sys)}}$\,ps was deduced, 
corresponding to a reduced transition probability of 
$B(E2; 5/2^-\!\rightarrow\!1/2^-)>2.6~e^2fm^4$ at the 1$\sigma$ level, with a best-fit value of $4.5~e^2fm^4$.
The measured transition strength is compared with state-of-the-art shell-model
and {\it ab initio} calculations, providing a stringent benchmark for shell-model Hamiltonians and effective charges
and modern {\it ab initio} descriptions of the neutron-rich calcium isotopes 
approaching the potential $N=40$ shell closure.
\end{abstract}







\end{frontmatter}



\section{Introduction}

Nuclear magic numbers provide the cornerstones of our understanding of nuclear structure. These numbers define the shell gaps and emergent phenomena of nuclear deformation, which appear in open-shell nuclei, giving rise to collective motion. However, their universality does not extend to exotic nuclei; some magic numbers vanish, while new ones appear away from the valley of stability. Nucleon-nucleon correlations are responsible for the modification of nuclear structure away from the valley of stability. Indeed, the isospin dependence of the monopole average of the central, tensor, and 2-body spin-orbit forces affects the effective single-particle energies and drives systems from magic, to deformed, and even to superfluid. 
Understanding and quantitatively describing properties of nuclei away from the valley of stability is at the forefront of nuclear structure science. 

The calcium isotopic chain has provided great input into our understanding of 
shell evolution, with new sub-shell gaps appearing at
$N=32$~\cite{Gade:2006:PRC} and $N=34$~\cite{Steppenbeck:2013:Nature}. 
Beyond \ts{54}Ca, the next potential shell closure is at $N=40$, 
marking the filling of the neutron $f_{5/2}$ orbital. 
The newly developed
$fp$ shell interaction, UFP-CA~\cite{Magilligan:2021:PRC}, fitted existing experimental data in calcium isotopes,
including the excitation energies up to \ts{54}Ca, and allows extrapolated predictions
out to \ts{60}Ca. It predicts that \ts{60}Ca is doubly magic at a similar level 
to \ts{68}Ni~\cite{Magilligan:2021:PRC}. In contrast, the first spectroscopy measurements of
\ts{56,58}Ca were recently performed at RIBF~\cite{Chen:2023:PLB}. The observed excitation energies, 
along with other experimental data on calcium isotopes, 
were used to refine the A3DA-m interaction in the $fpg_{9/2}d_{5/2}$ model space, which was named A3DA-t and predicted that 
\ts{60}Ca is not doubly magic and that the dripline 
of calcium isotopes extends to at least $N=50$~\cite{Chen:2023:PLB}. 
Additionally, low-lying states in \ts{55,57}Ca~\cite{Koiwai:2022:PLB} 
and \ts{62}Ti~\cite{Cortes:2020:PLB} were observed in the same 
experiment, also suggesting that the $N=40$ island of inversion extends
down to calcium isotopes.

There are still many open questions regarding the magicity of \ts{60}Ca \cite{Magilligan:2021:PRC,Nowacki:2021:PPNP,Chen:2023:PLB}.
In view of the difficulties of currently performing direct spectroscopy of \ts{60}Ca (with planned upgrades of facilities and detection systems targeting exactly this region), a critical next step is to benchmark theory against other observables which are sensitive to physics beyond what is relevant for excitation energies alone. In particular, $B(E2)$ strengths beyond $N=28$ ($Z=20$) have been intermittently studied experimentally, and new data are urgently needed to guide our  understanding of the Ca isotopic chain.
In this article we report on the first measurement of the lifetime of the first excited state ($5/2^-$) of $^{53}$Ca, and its $B(E2;5/2^-\rightarrow1/2^-)$ value, employing the HiCARI array at the RIBF facility and we compare this measurement with both shell-model calculations with effective charges, and {\it ab initio} calculations to test their description of this isotopic chain.


\section{Experimental setup}
\begin{figure}[!tb]
\centering
\includegraphics[width=0.48\textwidth]{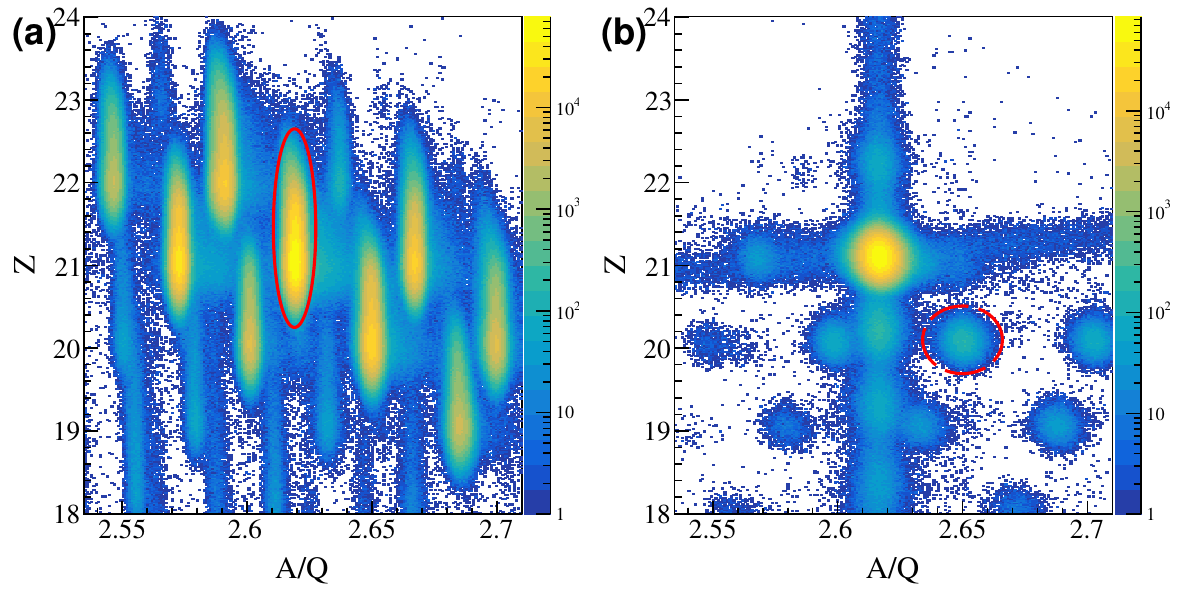}
\caption{\label{fig:pid}
(a) Incoming beam particle identification in the BigRIPS separator. The solid ellipse indicates the \ts{55}Sc. (b) Reaction residue particle identification in the ZeroDegree spectrometer for the \ts{55}Sc incoming beam. The dashed ellipse indicates the \ts{53}Ca.}
\end{figure}

The experiment was carried out at the Radioactive Isotope Beam Factory (RIBF) operated by the RIKEN Nishina Center and the Center for Nuclear Study (CNS), University of Tokyo.
Radioactive beams in the neutron-rich calcium region were produced via the fragmentation of 
a \ts{70}Zn primary beam on a \ts{9}Be target at an energy of 345\,\mevu. 
The fragments were selected using the BigRIPS separator~\cite{Kubo:2012:PTEP} and identified 
on an event-by-event basis through the TOF-B$\rho$-$\Delta$E measurements 
using the standard beam-line detectors described in Ref.~\cite{Fukuda:2013:NIMB}.
These measurements enabled unambiguous particle identification in terms of 
mass-to-charge ratio and atomic number, as shown in Fig.~\ref{fig:pid}(a).
A 1\,\gcm-thick carbon target and a 1.2\,\gcm-thick polyethylene (CH\tb{2}) target 
were placed at the F8 focal plane to induce the knockout reactions. 
The \ts{53}Ca nuclei in excited states were populated through the one-proton and one-neutron 
removal reaction (-1$p$-1$n$) from \ts{55}Sc projectiles.
The reaction residues were analyzed using the ZeroDegree spectrometer, 
applying the same method as in BigRIPS.
Figure~\ref{fig:pid}(b) shows the particle identification in the ZeroDegree spectrometer
gated on the incoming \ts{55}Sc beam.

The prompt \grays emitted following the knockout reactions were detected using 
the High-resolution Cluster Array at RIBF (HiCARI)~\cite{HiCARI}.
HiCARI is a hybrid array composed of multiple types of segmented high-purity germanium detectors.
In this experiment, six Miniball triple-cluster detectors (six-fold segmented)~\cite{miniball} 
were positioned at forward angles, covering $20^{\circ}$ -- $55^{\circ}$ relative to the beam axis.
Four clover detectors (four-fold segmented)~\cite{hua:2013:NSC} were mounted to cover $60^{\circ}$ -- $85^{\circ}$,
and two GRETINA-type tracking detector clusters~\cite{Weisshaar:2017:NIMA} were placed covering $50^{\circ}$ -- $90^{\circ}$.
The detector configuration is shown in Fig.~2 of Ref.~\cite{Acosta:2025:PRC}.
Each detector was calibrated individually using \ts{133}Ba, \ts{152}Eu, and \ts{60}Co \ghray sources.
The Doppler correction was performed by taking the point of highest energy deposit as 
the initial photon interaction. 
For the Miniball and clover detectors, the position of highest energy deposit segment -- pre-determined
from simulations as the average first interaction point for each segment -- was used 
to calculate the \ghray emission angle.
For the tracking detectors, 
pulse-shape analysis~\cite{Paschalis:2013:NIMA} was employed to determine 
the \ghray interaction positions within the crystal volume with a precision of a few millimeters.
Ultimately, the tracking detectors provided significantly superior energy resolution and 
peak-to-Compton ratios in the Doppler-corrected spectra compared with the Miniball and 
clover detectors, leading to higher sensitivity to the short lifetimes measured in this experiment. 
Therefore, only the tracking detectors were used in the subsequent analysis.

\section{Results}
For the \ghray analysis, Doppler corrections were applied using the mid-target velocities,
deduced event-by-event from the beam velocities measured with BigRIPS and ZeroDegree, 
accounting for the energy loss in the target material and beam-line detectors.
The \ghray emission point was assumed to be at the center of the target.
To improve the peak-to-Compton ratio, add-back analysis was performed by summing 
the detected energies within each cluster.

\begin{figure}[!tb]
\centering
\includegraphics[width=0.48\textwidth]{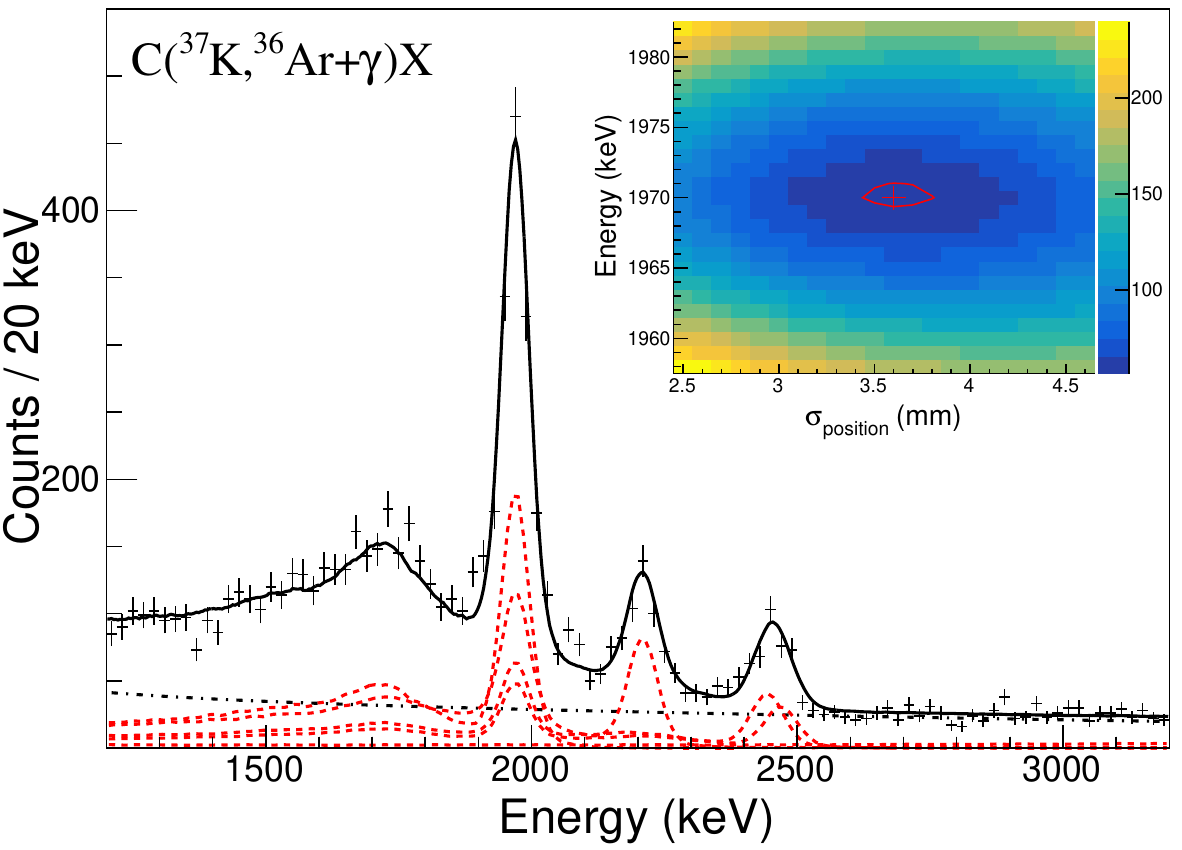}
\caption{\label{fig:spec36ArC}
Doppler-corrected \ghray energy spectrum of \ts{36}Ar, populated via one-proton removal from \ts{37}K on the carbon target, measured with the tracking detectors. The spectrum is fitted using simulated response functions with the final position resolution (red dashed line) on top of a double-exponential background (black dash-dotted line). The inset shows the $\chi^2$ surface for the \twotozero transition obtained by varying the transition energy and \ghray interaction position resolution in the simulation. The red cross marks the $\chi^2_\text{min}$, and the red contour indicates the 1$\sigma$ confidence region ($\chi^2_\text{min}+1$).}
\end{figure}

\begin{figure}[!tb]
\centering
\includegraphics[width=0.48\textwidth]{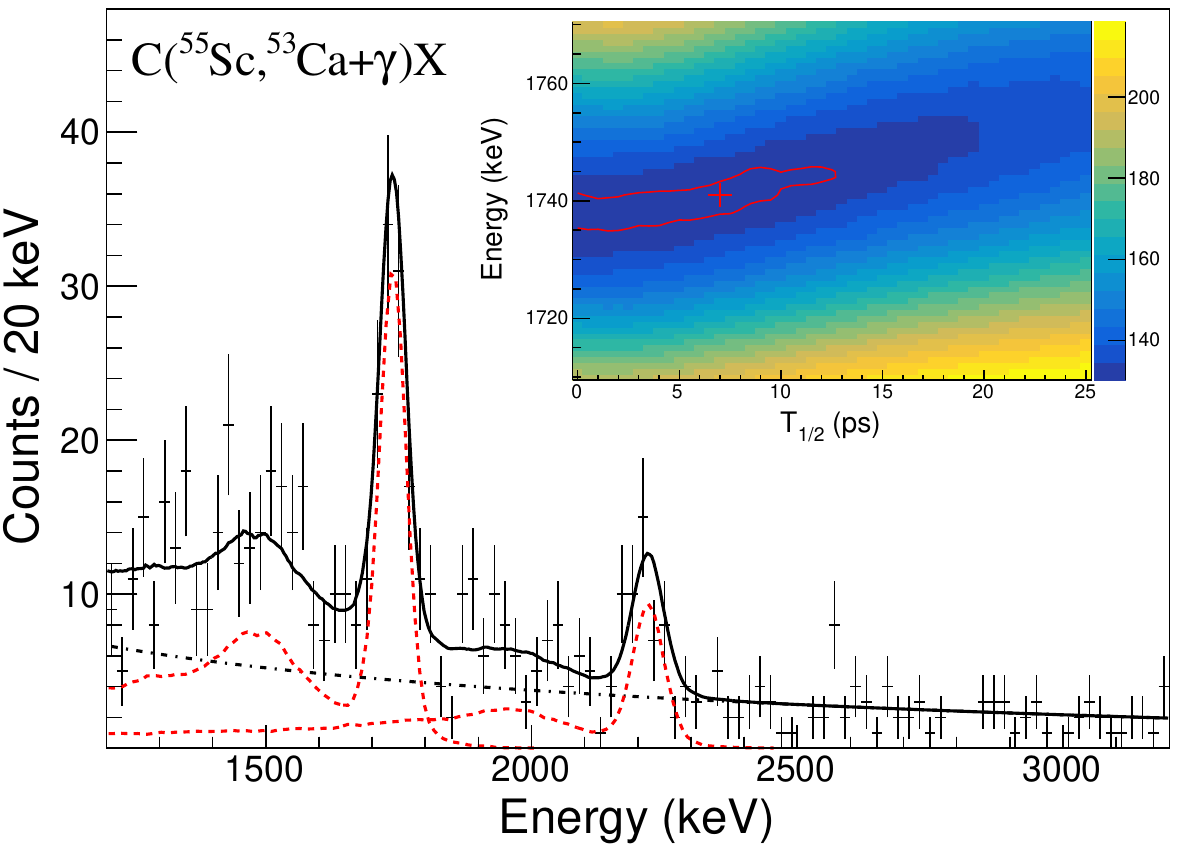}
\caption{\label{fig:spec53CaC}
The same as Fig.~\ref{fig:spec36ArC} but for \ts{53}Ca populated via one-proton and one-neutron removal from \ts{55}Sc. The inset shows the $\chi^2$ surface for the 1740-keV transition obtained by varying the transition energy and half-life in the simulation.}
\end{figure}

The resulting Doppler-corrected \ghray spectra for \ts{36}Ar (populated in -1$p$ from \ts{37}K) and \ts{53}Ca are shown in Fig.~\ref{fig:spec36ArC} and \ref{fig:spec53CaC}, respectively.
The observed spectral lineshapes are governed by three main factors: 
the intrinsic detector energy resolution, the velocity spread of the emitting nucleus,
and the uncertainty in the \ghray emission angle.
The lifetime affects both the velocity distribution at the time of \ghray emission 
and the angular uncertainty, thereby modifying the lineshape.
To extract lifetimes from the observed lineshapes, the experimental spectra were compared
with simulated response functions generated under different lifetime assumptions.
These simulations were performed using a customized Geant4~\cite{Agostinelli:2003:NIMA}
simulation package derived from the GRETINA simulation software~\cite{Riley:2021:NIMA}.
To reproduce the experimental lineshapes, the simulations incorporated 
the real target thickness and the measured beam profile, and energy resolutions 
obtained from source calibrations were implemented for each detector.
The effective position resolution of the tracking detectors -- critical for controlling 
the angular uncertainty -- was determined from the benchmark channel described below.
Doppler corrections were applied to the simulated data in the same manner as
for the experimental data to ensure consistency.

To benchmark the simulations, the spectra of \ts{36}Ar were analyzed first.
The low-lying excited states of \ts{36}Ar are well established, with both 
excitation energies and lifetimes accurately determined in previous studies~\cite{ENSDF}.
In this experiment, these excited states were populated via one-proton removal from 
a \ts{37}K beam, produced using the same experimental setup but 
with BigRIPS and ZeroDegree tuned to the neutron-deficient side.

The Doppler-corrected \ts{36}Ar spectrum obtained with the carbon target is shown in Fig.~\ref{fig:spec36ArC}.
Transitions from the \twoplus, \fourplus, \threeminus, and \twotwoplus states are clearly observed. 
According to the literature~\cite{ENSDF}, 
the \twoplus state (1970\,keV, $T_{1/2}=0.328$\,ps) decays directly to the ground state.
The \fourplus (4414\,keV, $T_{1/2}=0.076$\,ps) and \threeminus (4178\,keV, $T_{1/2}=2.3$\,ps) states
predominantly feed the \twoplus state, with negligible direct decay to the ground state.
The \twotwoplus state (4440\,keV, $T_{1/2}=0.076$\,ps) decays to 
both the \twoplus and ground states with a branching ratio of 36:64.
Simulations were therefore performed to generate response functions for cascade decay sequences of 
$2^+_1\rightarrow 0^+_\text{g.s.}$,
$4^+_1\rightarrow 2^+_1\rightarrow 0^+_\text{g.s.}$, 
$3^-_1\rightarrow 2^+_1\rightarrow 0^+_\text{g.s.}$, 
$2^+_2\rightarrow 2^+_1\rightarrow 0^+_\text{g.s.}$, and 
$2^+_2\rightarrow 0^+_\text{g.s.}$,
using the literature transition energies and excited state lifetimes.

The simulated response functions were combined with a double-exponential background to 
fit the experimental spectrum. The amplitudes of the response functions and 
the background parameters were treated as free parameters in the fit.
The branching ratio between the $2^+_2\rightarrow 2^+_1\rightarrow 0^+_\text{g.s.}$ 
and $2^+_2\rightarrow 0^+_\text{g.s.}$ decay paths was fixed to the literature value.

To determine the position resolution of the tracking detectors, the lineshape of 
the most intense transition, $2^+_1\rightarrow 0^+_\text{g.s.}$ at 1970\,keV, was investigated.
In this study, the $2^+_1\rightarrow 0^+_\text{g.s.}$ transition energy -- including those 
in the cascade decays -- and the position resolution were varied simultaneously in the simulations, 
while all other transition energies and all $T_{1/2}$ values were fixed to their literature values. 
The resulting $\chi^2$ surface as a function of the assumed transition energy 
and position resolution is shown in the inset of Fig.~\ref{fig:spec36ArC}. 
The minimum, $\chi^2_\text{min}$, was found at a transition energy of 1970\,keV -- consistent 
with the literature value -- and a position resolution of $\sigma=3.6$\,mm.
The extracted position resolution should be interpreted as an effective resolution 
that also accounts for uncertainties in the detector positions relative to the target.
The fit using the simulated response functions with this final resolution
is shown in Fig.~\ref{fig:spec36ArC}, demonstrating good reproduction of
the peak shapes and peak-to-Compton ratios observed in the experimental spectrum.
An analogous analysis was carried out for the \ts{36}Ar spectrum measured 
with the CH\tb{2} target, resulting in an effective position resolution of $\sigma=3.8$\,mm.

The data analysis of \ts{53}Ca followed the same procedure as that for \ts{36}Ar.
The Doppler-corrected \ts{53}Ca spectrum measured with the carbon target 
is shown in Fig.~\ref{fig:spec53CaC}. Excited states of \ts{53}Ca were populated
via one-proton and one-neutron removal from a \ts{55}Sc beam, 
the same reaction used in Ref.~\cite{Steppenbeck:2013:Nature}.
The observed spectrum is consistent with that measured using the DALI2 NaI(Tl) array 
in Ref.~\cite{Steppenbeck:2013:Nature}, but with significantly improved energy resolution, 
enabling lifetime sensitivity.

Two transitions are observed and were interpreted as decays from the $5/2^-$ 
and $3/2^-$ states directly to the $1/2^-$ ground state~\cite{Steppenbeck:2013:Nature}.
These spin-parity assignments were supported by momentum-distribution analysis
from knockout reactions in Ref.~\cite{Chen:2019:PRL}.
No additional bound excited states were observed in the high-statistic spectra
reported in Refs.~\cite{Steppenbeck:2013:Nature,Chen:2019:PRL}. As such,
feeding from higher-lying state was not considered in the present lifetime analysis.
Because of the limited statistics, the lifetime was extracted only for 
the stronger transition around 1.7\,MeV.
The transition around 2.2\,MeV has previously been measured in $\beta$-decay spectroscopy 
with an energy of 2220(1)\,keV~\cite{Perrot:2006:PRC}. Accordingly, this transition was 
fixed to the literature energy and assumed to have a negligible lifetime in simulations.
The transition around 1.7\,MeV has been measured only in the in-beam \ghray spectroscopy,
with reported energies of 1753(15)\,keV in Ref.~\cite{Steppenbeck:2013:Nature} and 
1738(17)\,keV in Ref.~\cite{Chen:2019:PRL}. Therefore, both the energy and lifetime 
of this transition were treated as free parameters and investigated simultaneously.
The tracking detector position resolutions determined from the \ts{36}Ar analysis
were applied in the corresponding simulations for each target.

Similar to the \ts{36}Ar case, the Doppler-corrected spectrum was fitted using simulated 
response functions combined with a double-exponential background.
For each combination of transition energy and lifetime, 
the corresponding $\chi^2$ value of the fit was evaluated.
The inset of Fig.~\ref{fig:spec53CaC} shows the resulting $\chi^2$ surface as a function of 
the assumed energy and half-life for the 1.7\,MeV transition.
The $\chi^2_\text{min}$ is marked with a red cross, and the red contour indicates 
the 1$\sigma$ confidence region ($\chi^2_\text{min}+1$).
The 1$\sigma$ region extends down to zero in lifetime, reflecting 
the limited sensitivity of the present setup to short lifetimes. 
The fit using the simulated response functions with the best-fit energy and lifetime is 
shown in Fig.~\ref{fig:spec53CaC}. An identical analysis of the \ts{53}Ca spectrum 
measured with the CH\tb{2} target yielded consistent results.
The extracted transition energies and lifetimes are summarized in Table~\ref{tab:BE2}, 
where only statistical uncertainties are included.
A weighted average of the results from both targets, using $1/\sigma^2$ as weights,
gives final values of $1741_{-5}^{+3}$\,keV and $\tau=11.3_{-11.3}^{+5.3}$\,ps.
The deduced transition energy agrees with the values reported in Ref.~\cite{Steppenbeck:2013:Nature,Chen:2019:PRL} within uncertainties.

\begin{table}[tb]
    \centering
    \caption{\label{tab:BE2}Summary of transition energies $E$, lifetime $\tau$, and reduced transition probabilities $B(E2)$ extracted for the $5/2^-\!\rightarrow\!1/2^-$ transition in \ts{53}Ca. Only statistical uncertainties are included.}
    \begin{tabular*}{\columnwidth}{@{\extracolsep{\fill}}lccc}
        \\[-10pt]\hline
        \hline
                   & $E$   &  $\tau$   & $B(E2; 5/2^-\!\rightarrow\!1/2^-)$ \\
                   & (keV) &  (ps)        & ($e^2fm^4$)\\
        \hline \\[-10pt]
        C target   & $1741_{-6}^{+4}$ & $10.1_{-10.1}^{+7.9}$ & $5.1_{-2.2}^{+\infty}$ \\[5pt]
        CH\tb{2} target & $1741_{-7}^{+4}$ & $12.3_{-12.3}^{+7.2}$ & $4.2_{-1.5}^{+\infty}$ \\[5pt]
        Weighted   & $1741_{-5}^{+3}$ & $11.3_{-11.3}^{+5.3}$ & $4.5_{-1.4}^{+\infty}$ \\
        \\[-10pt]\hline
        \hline
    \end{tabular*}        
\end{table}


Systematic uncertainties were also evaluated. The dominant contribution arises from 
the uncertainty in the position resolution.
An uncertainty of 0.3\,mm was adopted, based on the variation observed when 
extracting the position resolution for individual tracking clusters.
When translated into lifetime uncertainty, a smaller assumed position resolution 
leads to a longer extracted lifetime, and vice versa. 
Since the statistical lower limit of the lifetime is already at zero, 
only the upper systematic limit needs to be considered.
The same $\chi^2$ studies were therefore repeated assuming a smaller position resolution.
Under this assumption, the resulting 1$\sigma$ region extends to a lifetime that is longer by 3\,ps.
A systematic uncertainty of +3\,ps was thus adopted, yielding a final lifetime of
$\tau=11.3_{-11.3}^{+5.3\text{(stat)}+3\text{(sys)}}$\,ps, and a reduced transition probability of
$B(E2; 5/2^-\!\rightarrow\!1/2^-)>2.6~e^2fm^4$ at the 1$\sigma$ level, with a best-fit value of $4.5~e^2fm^4$.

\begin{table*}[tb]
\centering
\caption{\label{tab:exptheo}Reduced transition probabilities for the lowest $E2$ excitations in \ts{47--53}Ca isotopes. Experimental $B(E2)$ values are compared with shell-model calculations using KB3G, GXPF1Br, A3DA-t and UFP-CA Hamiltonians with standard $e_{n}=0.5$ effective charge, and UFP-CA calculations with microscopically-treated effective charges.}
    \begin{tabular*}{\textwidth}{@{\extracolsep{\fill}}ccccccccc}
        \\[-10pt]\hline
        \hline
                  &                          &        & $e_{n}=0.5$  & $e_{n}=0.5$ & $e_{n}=0.5$ & $e_{n}=0.5$  & micro $e_{n}$& \\
                  & $E2$ transition            & exp. $B(E2)$   & KB3G &GXPF1Br& A3DA-t & UFP-CA&UFP-CA& largest OBTD \\
        \hline\\[-10pt]
        \ts{47}Ca & $7/2^-\rightarrow 3/2^-$ & 2.0(1) & 0.4 & 1.6 & 1.5 & 2.0  & 1.0  & $p_{3/2}\rightarrow f_{7/2}$ \\
        \ts{48}Ca & $0^+\rightarrow 2^+$     & 95(9)  & 57.7   & 51.8 & 52.8 & 67.5   & 27.5 & $p_{3/2}\rightarrow f_{7/2}$ \\
        \ts{49}Ca & $3/2^-\rightarrow 7/2^-$ & 1.1(4) & 0.8  & 7.1 & 6.1 & 9.5  & 3.0  & $p_{3/2}\rightarrow f_{7/2}$ \\
        \ts{50}Ca & $0^+\rightarrow 2^+$  & 37.5(1.1) & 44.8   & 39.1 & 41.2 & 45.3   & 10.2 & $p_{3/2}\rightarrow p_{3/2}$ \\
        \ts{51}Ca & $3/2^-\rightarrow 7/2^-$ &    --    & 13.8 & 13.4 & 13.7 & 15.9 & 5.2  & $p_{3/2}\rightarrow f_{7/2}$ \\
        \ts{52}Ca & $0^+\rightarrow 2^+$     &     --   & 38.1   & 30.8 & 36.7 & 36.4 & 7.4  & $p_{1/2}\rightarrow p_{3/2}$ \\
        \ts{53}Ca & $1/2^-\rightarrow 5/2^-$ & $>7.8$ & 13.4 & 11.0 & 9.0 & 15.3 & 4.9  & $p_{1/2}\rightarrow f_{5/2}$ \\
        \\[-10pt]\hline
        \hline
    \end{tabular*}{}
\end{table*}

\section{Discussion}
\begin{figure}[!tb]
\centering
\includegraphics[width=0.48\textwidth]{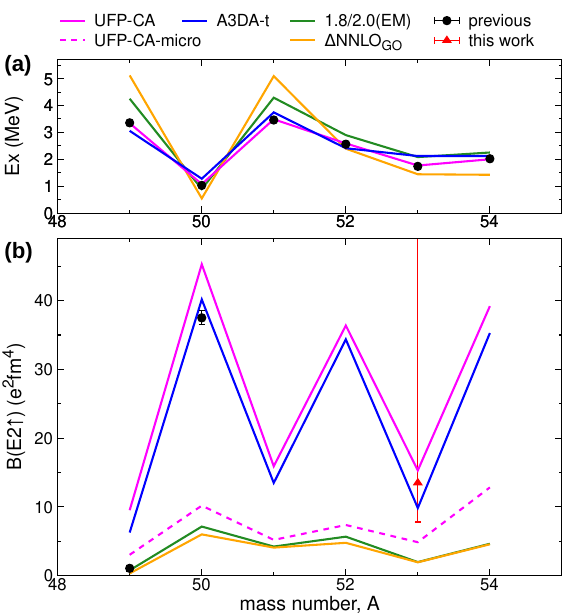}
\caption{\label{fig:BE2sys}
Systematic trend of excitation energies (a) and reduced transition probabilities (b) for the lowest $E2$ excitation in neutron-rich calcium isotopes as a function of mass number; $B(E2; 0^+ \rightarrow 2^+)$ for $^{50,52,54}$Ca,  $B(E2; 3/2^-\!\rightarrow\!7/2^-)$ for  $^{49,51}$Ca and $B(E2; 1/2^-\!\rightarrow\!5/2^-)$ for $^{53}$Ca . The experimental results are compared with theoretical predictions from {\it ab initio} approaches and shell-model calculations. Experimental data are taken from Refs.\cite{Montanari:2011:PLB,Valiente:2009:PRL} and from the present work.}
\end{figure}

In Fig.~\ref{fig:BE2sys} we present the  experimentally deduced reduced transition probabilities $B(E2\!\!\uparrow)$ 
for the lowest $E2$ excitations in neutron-rich calcium isotopes beyond the $N=28$ shell, previously measured only for \ts{49}Ca and \ts{50}Ca and this work delivering the first measurement for \ts{53}Ca.
For \ts{49}Ca, the value is taken from Ref.~\cite{Montanari:2011:PLB} for 
the $3/2^-_\text{g.s.}\!\rightarrow\!7/2^-$ excitation, while the value for 
\ts{50}Ca is adopted from Ref.~\cite{Valiente:2009:PRL} for 
the $0^+_\text{g.s.}\!\rightarrow\!2^+$ excitation.
Both literature values were deduced from lifetime measurements (using 
the Recoil Distance Doppler Shift method).
The experimental results are compared with shell-model calculations employing 
the UFP-CA~\cite{Magilligan:2021:PRC} and A3DA-t~\cite{Chen:2023:PLB} interactions (with standard effective charges),
as well as {\it ab initio} calculations using two NN+3N forces, 1.8/2.0(EM)~\cite{PhysRevC.83.031301,PhysRevC.96.014303} and  $\Delta$NNLO$_{\text{GO}}$~\cite{PhysRevC.102.054301}, derived from chiral EFT.
In addition, UFP-CA calculations with 
microscopically-treated effective charges (UFP-CA-micro) are shown.
All calculations provide 
reasonable descriptions of the excitation energies (Fig.~\ref{fig:BE2sys}(a)).
A more interesting picture emerges for the transition strengths 
shown in Fig.~\ref{fig:BE2sys}(b), which we will discuss separately for each theoretical approach (shell model vs {\it ab initio}).


\subsection{Shell-model calculations and effective charges}

For shell-model calculations, the $B(E2)$ depends on the effective-charge parameters $e_p$ and $e_n$. The effective charges account for the renormalization of the proton and neutron components of the $E2$ matrix elements within the shell-model basis of a major shell due to admixtures of the 1p-1h, $\Delta N=2$ proton excitations. 
Effective charges are not expected to be constant as a function of increasing asymmetry ($N$, $Z$), but to have an approximate $1/A$ dependence, as understood by the macroscopic model of Bohr and Mottelson \cite{BohrMottelson} and exemplified in  \cite{Petri:2011:PRL}, and by the microscopic calculations of Sagawa {\it et al.}~\cite{Sagawa:2004:PRC} and Longfellow {\it et al.}~\cite{Longfellow:2021:PRC}.

Experimental $B(E2)$ values for the neutron-rich calcium isotopes
are compared to calculations in Table~\ref{tab:exptheo}. The theory includes those from
two standard Hamiltonians for the $pf$ model space, KB3G~\cite{Poves:2001:NPA}
and GXPF1Br~\cite{Honma:2005:EPJA,Steppenbeck:2013:Nature}, together with the more recent A3DA-t~\cite{Chen:2023:PLB} and
UFP-CA~\cite{Magilligan:2021:PRC} Hamiltonians. The $B(E2)$ are obtained with the
standard value of $e_{n}=0.5$ for the neutron effective charge
from \cite{Honma:2005:EPJA}. We use harmonic radial wavefunctions
with $\hbar\omega = 45A^{-1/3}-25A^{-2/3}$.
For all of these Hamiltonians the one-body
transition densities (OBTD) are dominated by one
term given in the last column of Table~\ref{tab:exptheo}.
The calculated results for \ts{50,51,52,53}Ca
are similar for all three Hamiltonians and are in
good agreement with experiment. The agreement for \ts{50}Ca
is improved if the slightly smaller effective charge
from \cite{Ogunbeku:2025:PRL} of $e_{n}=0.45$ is used.

For \ts{42,44}Ca, the $B(E2)$ obtained in the $pf$
model space are an order of magnitude smaller than
experiment (see Fig.~17 in \cite{Longfellow:2021:PRC}).
One must include $sd$ to $pf$ excitations for
the low-lying intruder state and their mixing with
the $pf$ configurations~\cite{Brown:2022:PRC}.
The intruder states
move up in energy as the neutron number increases,
so that the wavefunctions above \ts{46}Ca have
more pure $pf$ configurations. The factor of two
larger experimental $B(E2)$ value for \ts{48}Ca compared the
$pf$ calculations may be due to $sd$ to $pf$ admixtures.

For \ts{47,49}Ca the calculated $B(E2)$ are relatively
small and depend more strongly on the Hamiltonian.
In the $pf$ model space the wavefunctions are dominated by the
simple configurations of
\begin{equation}
    \begin{aligned}
        [(0f_{7/2})^{7}]&\,\,\,(7/2^{-}) \rightarrow \\
        &[(0f_{7/2})^{6},J_{f}]\otimes[(1p_{3/2})]\,\,\,(3/2^{-})
    \end{aligned}
\end{equation}
for \ts{47}Ca and
\begin{equation}
    \begin{aligned}
        [(0f_{7/2})^{8}]&\otimes[1p_{3/2}]\,\,\,(3/2^{-}) \rightarrow \\
        &[(0f_{7/2})^{7},7/2^{-}]\otimes[(1p_{3/2})^{2},J_{p}]\,\,\,(7/2^{-})
    \end{aligned}
\end{equation}
for \ts{49}Ca. The largest calculated $B(E2)$
values come from the configurations with $J_{f}=J_{p}=0$.
However, the $B(E2)$ become reduced if there is
admixture and
interference with the components with $J_{f}=2$ for \ts{47}Ca and $J_{p}=2$
for \ts{49}Ca.
In this regard, the mixing obtained with
the KB3G Hamiltonian for \ts{49}Ca is the only
one that gives agreement with experiment.

\begin{figure}[!tb]
\centering
\includegraphics[width=0.48\textwidth]{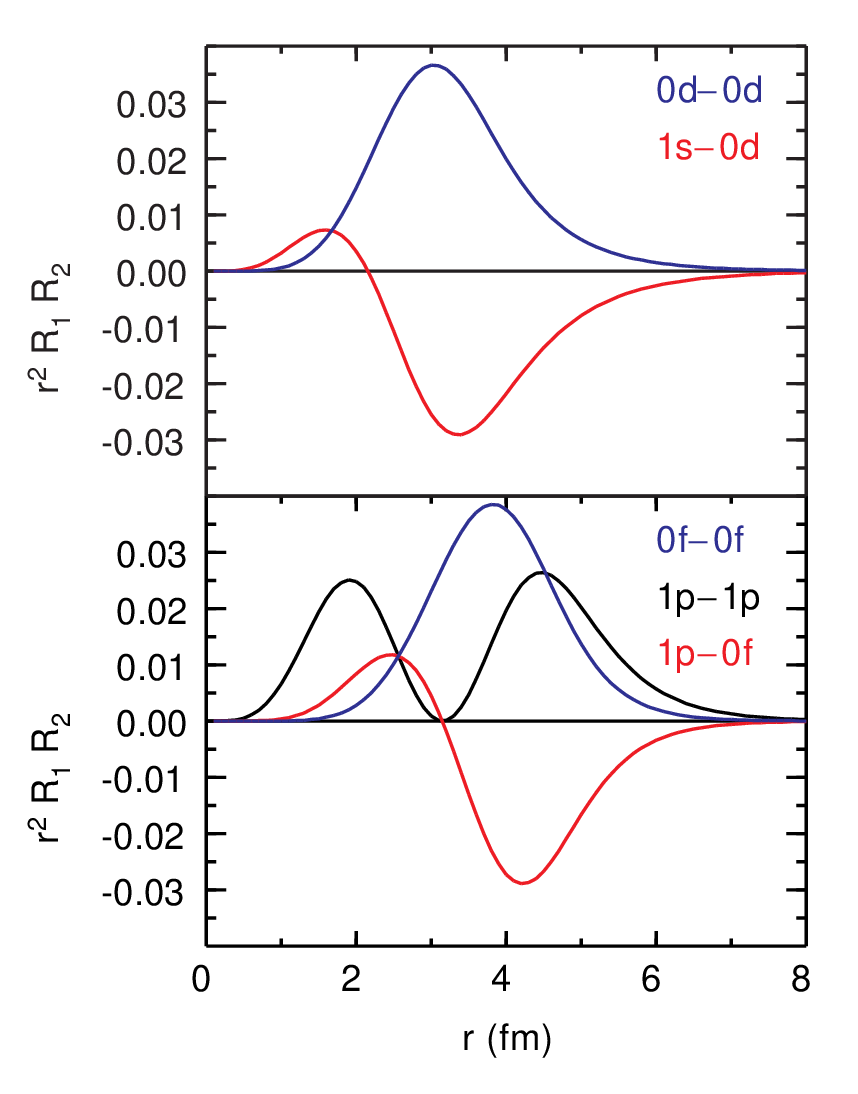}
\caption{\label{fig:w}
Radial parts of the valence single-particle $E2$ transition densities 
for the neutrons in the $sd$ shell for $^{22}$O (top), and for neutrons
in the $pf$ shell for $^{48}$Ca (bottom).}
\end{figure}

The $pf$ shell description for \ts{47}Ca is analogous to the
the $sd$ description of the $5/2^+$ to $1/2^+$ transition
discussed in \cite{Heil:2020:PLB} with dominant configurations
\begin{equation}
    \begin{aligned}
        [(0d_{5/2})^{5}]&\,\,\,(5/2^{+}) \rightarrow \\
        &[(0d_{5/2})^{4},J_{d}=0]\otimes[(1s_{1/2})]\,\,\,(1/2^{+})
    \end{aligned}
\end{equation}
Note that we can only have $J_{d}=0$ for \ts{21}O.
In \cite{Heil:2020:PLB} it was noted that the $sd$ calculation
with the USDB Hamiltonian~\cite{Brown:2006:PRC} and with the $sd$
model space effective charge of $e_{n}=0.45$~\cite{Richter:2008:PRC} gave
a $B(E2)$ that was about a factor of two larger than
experiment for this $5/2^+$ to $1/2^+$ transition in \ts{21}O. This motivated discussion about
the microscopic origin of the effective charge.
We used the first-order model of \cite{Brown:1977:NPA}
where the effective charge is obtained from
an integral involving the valence transition density times the
collective transition
density for the giant quadrupole excitation
that peaks near the nuclear surface.
It was noted in this model that the effective charge for
the $0d_{5/2}$ to $1s_{1/2}$ transition
was reduced due to the crossing node in the
$0d-1s$ valence transition density shown in Fig.~\ref{fig:w}.
For $^{50}$Ca and $^{52}$Ca, the integral involving the $p$
to $p$ valence transition has a minimum at 3~fm as shown in Fig.~\ref{fig:w}.
For the other cases in Table~\ref{tab:exptheo} involving $p$ to $f$, the valence
transition density crosses zero near 3~fm.
This results in a small effective charge (see Table~3 in \cite{Longfellow:2021:PRC}).
The $B(E2)$ obtained with the microscopic model are given
in Table~\ref{tab:exptheo}. The resulting $B(E2)$ for \ts{50,53}Ca
are about a factor of four smaller than experiment.
This additional information indicates that the
first-order model is inadequate.
This leaves open the question
as to why the experimental $B(E2)$ for \ts{21}O is a factor
of two smaller than the calculation.

 \subsection{\textit{Ab initio} calculations}

 {\it Ab initio} theory has seen rapid recent progress for descriptions of ground-state energies and radii as well as the excitation spectrum. 
 First microscopic calculations~\cite{Hagen:2012:PRL,Holt:2013vqa,Holt:2014:PRC} have also highlighted the calcium isotopes as a key isotopic chain to test the role of 
 3N forces in the quest for a microscopic description of the atomic 
 nucleus using realistic interactions originating from chiral EFTs. 
 {\it Ab initio} calculations of masses,  charge radii, and low-lying spectroscopy have also been performed for neutron-rich calcium isotopes~\cite{Wienholtz:2013:Nature,GarciaRuiz:2016:NaturePhy,hu2025:texas,Chen:2023:PLB,Koiwai:2022:PLB,Cortes:2020:PLB}. 
 The critical next step is the description of electromagnetic observables, addressing the full spectroscopy of low-lying states. 

\begin{figure}[!tb]
\centering
\includegraphics[width=0.48\textwidth]{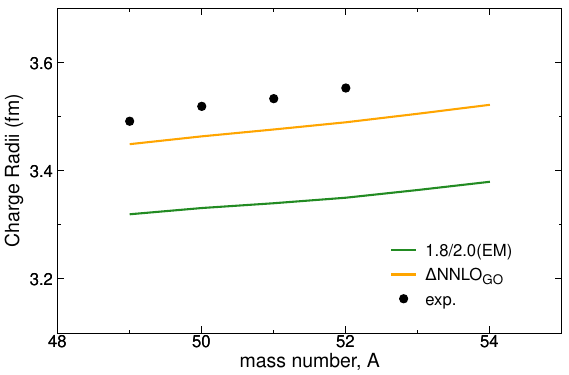}
\caption{\label{fig:Rch}
Charge radii in neutron-rich calcium isotopes as a function of mass number. The experimental results \cite{GarciaRuiz:2016:NaturePhy, PhysRevC.105.034320} are compared with \emph{ab initio} VS-IMSRG predictions utilizing the 1.8/2.0(EM) and $\Delta$NNLO$_{\text{GO}}$ interactions.}
\end{figure}

 The valence-space in-medium similarity renormalization group 
 (VS-IMSRG)~\cite{tsukiyama2011,hergert2016,stroberg2017,Stroberg:2019mxo,hu2022} were performed to produce {\it ab initio} $fp$-shell valence-space Hamiltonians and  consistently evolved $E2$ operators~\cite{Parzuchowski:2017wcq}, based on NN+3N forces, 1.8/2.0(EM)~\cite{PhysRevC.83.031301,PhysRevC.96.014303} and  $\Delta$NNLO$_{\text{GO}}$~\cite{PhysRevC.102.054301}, derived from chiral EFT. 
 For $E2$ observables, the major source of theory uncertainties in this case is the missing higher-order collective excitations~\cite{Henderson:2017dqc,Stroberg:2022ltv}.
 Already in its leading-order or single-particle form, the $E2$ operator probes the collective quadrupole deformation  of the nuclear states as well as the long-range behaviour of the wave functions. 
 This leads to a significantly slower many-body convergence compared to, e.g., energies or radii, when calculations are performed within a spherical basis. Therefore, many-body truncations are the dominant limitation in {\it ab initio} VS-IMSRG calculations of $E2$ observables. 
 Another source of uncertainty can in principle be correlated with the calculated charge radii of these nuclei. The $E2$ operator is proportional to $r^2$ ($r$ being the point proton radius), and any underprediction in the charge radii could be expected to drive the suppression of $B(E2)$ values. Figure~\ref{fig:Rch} shows the experimental charge radii in neutron-rich Ca isotopes and how they compare with the VS-IMSRG calculations. In particular, we note that those using the $\Delta$NNLO$_{\text{GO}}$ interaction are reproducing relatively well the charge radii, and therefore we expect this should not be the major contributor to the underestimation of the $B(E2)$ values. 

 For the odd-mass cases the agreement of the VS-IMSRG calculations with experiment is generally good, as was similarly seen in \ts{21}O~\cite{Heil:2020:PLB} and \ts{49}Ca~\cite{Montanari:2011:PLB}. In contrast to shell-model calculations where the $B(E2)$ strength depends on the effective charges, the latter are not relevant to the VS-IMSRG calculations, since the effective
operators are calculated consistently, and therefore effective charges are not needed. It is interesting to see that VS-IMSRG (which generally underestimates more collective transitions in even-even isotopes) are closer to the UFP-CA calculations with microscopically-treated effective charges. 

The investigation of associated shell model effective charges -- and how they arise from many-body mechanisms -- is being approached with self-consistent Green’s function (SCGF) theory~\cite{Raimondi:2019:PRC}. In this framework, the knowledge of electromagnetic transition strength for neutron-rich  calcium isotopes is pivotal to test and improve the SCGF calculations.

\section{Conclusion}
In summary, the first high-resolution in-beam $\gamma$-ray spectroscopy of \ts{53}Ca
was performed at the RIBF using the HiCARI array.
The lifetime of the first excited $5/2^-$ state was determined 
for the first time through a Doppler line-shape analysis.
The resulting $B(E2)$ value is well reproduced by shell-model
calculations using the standard neutron effective charge $e_{n}=0.5$;
however, calculations employing microscopically-treated effective charges 
and modern {\it ab initio} approaches substantially underestimate the observed strength.
The present result adds to the electromagnetic transition data in the neutron-rich Ca isotopes,
which provide a stringent benchmark for theoretical descriptions of
electromagnetic transitions in neutron-rich nuclei. 
Further experimental investigations, such as systematic measurements of 
transition probabilities in neutron-rich calcium isotopes via Coulomb excitation
will provide a more comprehensive picture of shell evolution in this region. 

\section*{Acknowledgements}

We thank Augusto~O.~Macchiavelli for insightful discussions.
This work was carried out at the RIBF operated by RIKEN Nishina Center and 
Center for Nuclear Study in the University of Tokyo and we thank the staff for providing
stable beams with high intensities to the experiment. 
We acknowledge the Miniball Collaboration for the loan of the Miniball detectors and 
the supporting frame. The Research Center for Nuclear Physics in Japan, 
the Institute of Modern Physics in China, the Lawrence Berkeley National Laboratory, 
and the Argonne National Laboratory in the USA are acknowledged for 
the loan of detectors and data acquisition electronics.
The HiCARI campaign was supported by the JSPS KAKENHI of Japan under Grants No. Kiban-A
JP19H00670 and No. JP19H01914.
This work was supported by the Royal Society and the UK STFC under grant numbers ST/P003885/1, ST/V001035/1, and ST/Y000285/1.
B.~A.~B. acknowledges support from NSF Grant No. PHY-2110365. J.~D.~H. acknowledges support from NSERC
under grant SAPIN-2024-0003, and the Arthur B. McDonald Canadian Astroparticle Physics Research Institute. 
T.~M. acknowledges support from JST ERATO Grant No. JPMJER2304, Japan and KAKENHI (25K07294, 25K00995, 25K07330, 26H01394).
The VS-IMSRG calculations were performed using the \texttt{imsrg++}~\cite{code-imsrg} code. Computations were performed with an allocation of computing resources on Cedar and Fir at WestGrid and the Digital Research Alliance of Canada. 
B.~M. acknowledges support from Institute for Basic Science (Grant Nos. IBS-R031-Y1 and IBS-R031-D1) and National Research Foundation (Grant No. 2019R1A6A3A03031564) of the Republic of Korea.
This work is partially supported by JSPS KAKENHI (Grant Nos. 19H00679, 19H01914 and 24H00239)).
K.~W. acknowledges support from the European Research Council (ERC) under the European Union’s Horizon 2020 research and innovation programme (grant agreement No 101001561) and the Deutsche Forschungsgemeinschaft (DFG, German Research Foundation) under Contract No. SFB 1245 (Project ID No. 79384907). 
H.~L.~C., P.~F. and C.~C. acknowledge support from U.S. Department of Energy, Office of Science, Office of Nuclear Physics, under Contract No. DE-AC02-05CH11231 (LBNL).




\bibliographystyle{elsarticle-num}
\bibliography{bibliography}






\end{document}